\documentclass[aps,eqsecnum,preprintnumbers]{revtex4}
\usepackage{axodraw}
\newcommand{\beq}{\begin{equation}}
\newcommand{\eeq}{\end{equation}}
\newcommand{\beqa}{\begin{eqnarray}}
\newcommand{\eeqa}{\end{eqnarray}}
\newcommand{\lslash}[1]{#1\llap/}

\newcommand{\Eq}[1]{Eq.\ (\ref{#1})}
\newcommand{\Eqs}[2]{Eqs.\ (\ref{#1}) and (\ref{#2})}

\newcommand{\Ref}[1]{Ref.\ \cite{#1}}
\newcommand{\Tr}{\mbox{Tr}\,}

\begin{document}

\preprint{hep-ph/0502227}

\title{Electromagnetic effects of neutrinos in an electron gas}

\author{Jos\'e F. Nieves}
\affiliation{Laboratory of Theoretical Physics,
Department of Physics, P.O. Box 23343,
University of Puerto Rico
R\'{\i}o Piedras, Puerto Rico 00931-3343}

\author{Sarira Sahu}
\affiliation{Instituto de Ciencias Nucleares, Universidad Nacional
Aut\'onoma de M\'exico, Circuito Exterior, C. U.,
A. Postal 70-543, 04510 Mexico DF, Mexico}

\begin{abstract}
We study the electromagnetic properties of a system 
that consists of an electron background and a neutrino gas that 
may be moving or at rest, as a whole, relative to the background. 
The photon self-energy for this system is characterized
by the usual transverse and longitudinal polarization functions, 
and two additional ones which are the focus of our calculations,
that give rise to birefringence and anisotropic effects in the
photon dispersion relations. Expressions for them
are obtained, which depend on the neutrino number densities and involve
momentum integrals over the electron distribution functions,
and are valid for any value of the photon momentum and general
conditions of the electron gas. Those expressions are evaluated
explicitly for several special cases and approximations which are
generally useful in astrophysical and cosmological settings.
Besides studying the photon dispersion relations, 
we consider the macroscopic electrodynamic equations for this system,
which involve the standard dielectric and permeability constants
plus two additional ones related to the photon self-energy functions.
As an illustration, the equations are used to discuss the
evolution of a magnetic field perturbation in such a medium. This
particular phenomena has also been considered in a recent work by
Semikoz and Sokoloff as a mechanism for the generation of 
large-scale magnetic fields in the Early Universe as a consequence 
of the neutrino-plasma interactions, and allows us to establish contact 
with a specific application in a well defined context, 
with a broader scope and from a very different point of view.
\end{abstract}

\maketitle

%
%
\section{Introduction and Summary}

This work is concerned with the electromagnetic properties
of a medium that consists of a matter background, such as an electron
plasma, and a neutrino gas that moves, as a whole, relative to the
matter background. Technically, the quantity of interest to us is the
photon self-energy, from which the dispersion relations of the
photon modes that propagate in the medium can be obtained, and from which
other macroscopic quantities of physical interest can be determined.

Some aspects of this composite system were studied in \Ref{nievesinstabilities}
using the methods of real-time finite temperature 
field theory (FTFT)\footnote{%
For reviews, see e.g., Refs.\ \cite{landsman,kapusta,bellac}.} which, from a 
modern point of view, provides a natural setting for studying
the problems related to the propagation of photons in a medium.
Largely stimulated by the work of Weldon\cite{weldon1,weldon2,weldon3},
a convenient technique employed in
FTFT is to carry out the calculations in a manifestly covariant
form. As is by now familiar, this is implemented by introducing
the velocity four-vector $u^\mu$ of the medium, in terms of which
the thermal propagators are written in a covariant form. In this
way, covariance is maintained, but quantities such as the photon
self-energy depend on the vector $u^\mu$ in addition to the 
kinematic momentum variables of the problem. Generally, for
practical purposes the vector $u^\mu$ is set to $(1,\vec 0)$ in the end,
which is equivalent to having carried out the calculation in the
rest frame of the medium from the beginning, and this is usually
the relevant physical situation. 

However, as noted in \Ref{nievesinstabilities},
the system we are considering provides 
an example for a novel application of FTFT. 
A distinctive feature of this system
is that the matter background on one hand, and the neutrino gas on the other,
each is characterized by its own velocity four-vector. 
Thus, if we denote by $u^\mu$ the velocity
four-vector of the matter background, and by $v^\mu$ the corresponding one
for the neutrino gas, then we can take the 
matter background to be at rest, so that
\beq
u^\mu = (1, \vec 0) \,,
\eeq
but we must keep
\beq
v^\mu = (v^0, \vec V) \,.
\eeq
Therefore, the physical quantities such as the photon self-energy 
depend on the momentum variables and $u^\mu$ as usual,
and in addition on the vector $v^\mu$. This additional dependence
can have physical effects that cannot be produced by the
stationary background alone.

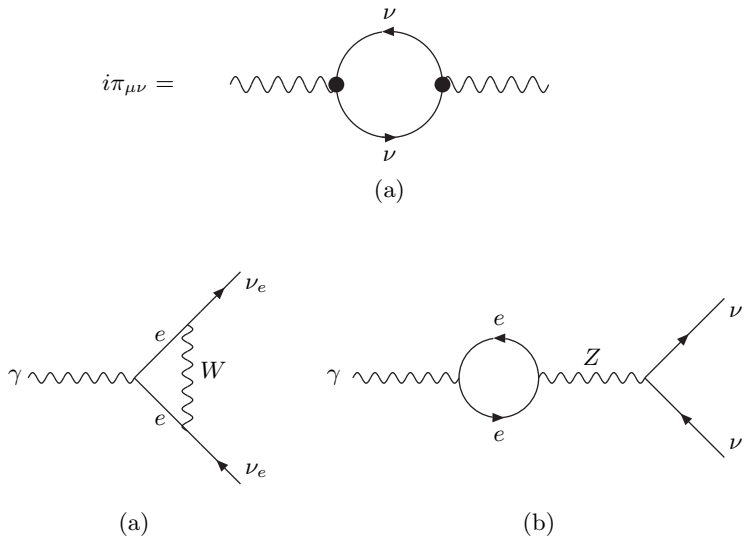
\begin{figure}
\begin{center}
%
%
\begin{picture}(200,120)(-100,-60)
\Photon(-60,0)(-20,0){3}{5}
\Photon(20,0)(60,0){3}{5}
\Vertex(-20,0){3}
\Vertex(20,0){3}
\LongArrowArc(0,0)(20,-98,98)
\LongArrowArc(0,0)(20,98,278)
\Text(0,25)[b]{$\nu$}
\Text(0,-25)[t]{$\nu$}
\Text(-95,0)[]{$i\pi_{\mu\nu} =$}
\Text(0,-45)[b]{(a)}
\end{picture}
\\
%
%
\begin{picture}(100,100)(-50,-50)
\Text(0,-60)[cb]{(a)}
\Photon(-40,0)(0,0){2}{6}
\Text(-45,0)[c]{$\gamma$}
\Line(0,0)(25,25)
\Text(12,14)[rb]{$e$}
\ArrowLine(25,25)(40,40)
\Text(42,35)[lc]{$\nu_e$}
\ArrowLine(40,-40)(25,-25)
\Line(25,-25)(0,0)
\Text(12,-14)[rt]{$e$}
\Text(42,-35)[lc]{$\nu_e$}
\Photon(20,20)(20,-20){2}{6}
\Text(25,0)[lb]{$W$}
\end{picture}
%
%
\begin{picture}(190,100)(-140,-50)
\Text(-40,-60)[cb]{(b)}
\Photon(-110,0)(-70,0){2}{6}
\Text(-115,0)[cr]{$\gamma$}
\LongArrowArc(-55,0)(15,-98,98)
\LongArrowArc(-55,0)(15,98,278)
\Text(-55,20)[b]{$e$}
\Text(-55,-20)[t]{$e$}
\Photon(-40,0)(0,0){2}{6}
\Text(-20,5)[b]{$Z$}
\ArrowLine(0,0)(30,30)
\Text(32,25)[lc]{$\nu$}
\ArrowLine(30,-30)(0,0)
\Text(32,-25)[lc]{$\nu$}
\end{picture}
\end{center}
\caption[]{A class of diagrams that contribute to the photon self-energy
in the presence of a neutrino background.
The filled circle in diagram (a) represents the neutrino electromagnetic 
vertex that is induced by the interactions with the electron background\cite{%
oraevsky85,oraevsky87,semikoz87a,semikoz87b,np1,sawyer}, 
which can be determined by calculating the off-shell amplitude corresponding
to diagrams (b) and (c)\cite{dnp1}.
\label{fig:pipnu}
}
\end{figure}

The focus of attention in \Ref{nievesinstabilities} was the effect
that the collective neutrino-plasma interactions could have on the stability
of such systems\cite{silvaetal,bento}, in analogy with the
\emph{stream instabilities} that are familiar in plasma
physics research, examples of which are discussed in many 
textbooks\cite{ishimaru}.
The calculation was focused on the contribution to the
photon self-energy from the set of diagrams represented in 
Fig. \ref{fig:pipnu}.
%
%

On the other hand, it was recognized some time 
ago\cite{nppip,nppipE,mohantynp,repko} 
that the presence of neutrinos induces 
\emph{birefringence} effects in the medium, similar to those that exist
in materials that exhibit \emph{natural optical activity}\cite{npactvityconst}.
These effects are a consequence of a new term, denoted by $\pi_P$ 
in \Ref{nppip}, that shows up
in the photon self-energy, which is proportional to the
neutrino-antineutrino asymmetry and is odd under various
discrete space-time symmetries. However, the previous calculations
of these effects have taken into account only the neutrino background.
In particular, for the purpose of determining $\pi_P$ in \Ref{mohantynp},
the background was taken to consist only of neutrinos and antineutrinos.
Therefore, for all the other particles the vacuum propagator
was used in the calculation.

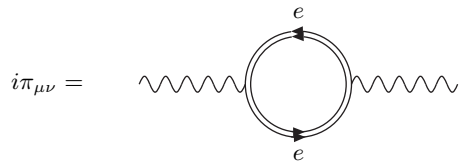
\begin{figure}
\begin{center}
%
%
\begin{picture}(200,120)(-100,-60)
\Photon(-60,0)(-20,0){3}{5}
\Photon(20,0)(60,0){3}{5}
\LongArrowArc(0,0)(20,-98,98)
\LongArrowArc(0,0)(20,98,278)
\LongArrowArc(0,0)(18,-98,98)
\LongArrowArc(0,0)(18,98,278)
\Text(0,25)[b]{$e$}
\Text(0,-25)[t]{$e$}
\Text(-95,0)[]{$i\pi_{\mu\nu} =$}
\end{picture}
\end{center}
\caption[]{A class of diagrams that contribute to the photon self-energy
in the presence of a neutrino background. The double electron line 
represents the thermal electron propagator, with the effects
of the neutrino gas taken into account. 
\label{fig:pipenu}
}
\end{figure}

\begin{figure}
\begin{center}
%
%
\begin{picture}(180,130)(-90,-30)
\Text(0,-30)[c]{(a)}
\ArrowLine(80,0)(40,0)
\Text(60,-10)[c]{$e(p)$}
\ArrowLine(40,0)(-40,0)
\Text(0,-10)[c]{$\nu_e(k)$}
\ArrowLine(-40,0)(-80,0)
\Text(-60,-10)[cr]{$e(p)$}
\PhotonArc(0,0)(40,0,180){4}{6.5}
\Text(0,50)[cb]{$W(p - k)$}
\end{picture}
%
%
\begin{picture}(100,100)(-50,-30)
\Text(0,-30)[c]{(b)}
\ArrowLine(40,0)(0,0)
\Text(35,-10)[cr]{$e(p)$}
\ArrowLine(0,0)(-40,0)
\Text(-35,-10)[cl]{$e(p)$}
\Photon(0,0)(0,35){2}{6}
\Text(-4,20)[r]{$Z$}
\ArrowArc(0,55)(20,-90,270)
\Text(0,85)[b]{$\nu_\alpha(q)$}
\end{picture}
\caption[]{One-loop diagrams for the electron self-energy 
in a neutrino gas.
\label{fig:sigmae}
}
\end{center}
\end{figure}
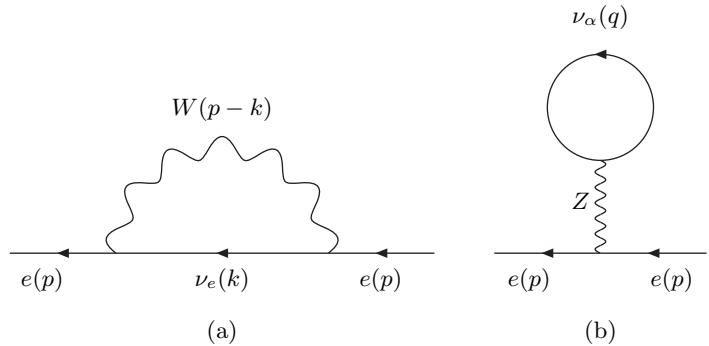

Here we note that, in the presence of neutrinos,
there is an important contribution to the photon self-energy
which is proportional to both the neutrino asymmetry and the electron
number density. It is best specified by referring to
the diagram depicted in Fig. \ref{fig:pipenu}.
%
%
As indicated by the double line in this diagram,
the electron propagator to be used in 
the calculation is the \emph{dressed} propagator with the 
correct electron self-energy in the medium. The electron self-energy includes 
the effect of the neutrino gas, as represented
by the diagrams shown in Fig.\ \ref{fig:sigmae}.
%
%
As we show, this contribution to the photon self-energy is of no consequence
for the instability issues, but it yields a novel contribution 
to the optical activity of the system, that can lead to
interesting and significant physical effects. In particular,
we note that the contribution to $\pi_P$ calculated in \Ref{mohantynp}
is proportional to the photon momentum squared $q^2$. In contrast,
the contribution that is determined here has no such term, and 
it is the only one that survives in the $q\rightarrow 0$ limit. This
feature leads to effects that can manifest themselves at a 
macroscopic level, in the static and long wavelength electromagnetic regime.

In this work we compute the contribution to the photon self-energy
that is proportional to both the neutrino-antineutrino asymmetry and
the electron density, and consider the consequences for the propagation
of photons and for the electromagnetic properties of the medium.
The calculation is based on the application of FTFT to calculate the
photon self-energy diagram shown in Fig. \ref{fig:pipenu}, using the
electron propagator that includes the effect of the neutrino-electron
interactions. The implicit assumption
is that, in its own rest frame, the neutrino gas
has a momentum distribution function that is
parametrized in the usual way. The presence of the neutrinos gives 
rise to two additional polarization functions
besides the ordinary longitudinal and transverse polarization functions
the photon self-energy, that we denote by 
$\pi_P$ and $\pi^{\prime}_P$. A non-zero value of 
$\pi_P$ by itself leads to optical activity
effects, while $\pi^\prime_p$ induces anisotropic effects. 
Aside from the implications for the
photon dispersion relations, the results for the photon self-energy can be
interpreted in terms of the macroscopic electromagnetic properties
of the system. In particular, the effects due to 
$\pi_P$ and $\pi^{\prime}_P$ remain finite in the limit $q \rightarrow 0$,
and therefore they can manifest themselves in macroscopic effects in
the long wavelength and static regimes.
As a specific application, we consider the evolution of a 
magnetic field perturbation in the system,
and we arrive at an equation for the dynamics of the magnetic field
that has been suggested by Semikoz and Sokoloff\cite{semikozalpha},
in the context of a mechanism for the generation of 
large-scale magnetic fields 
in the Early Universe as a consequence of the neutrino-plasma interactions. 
Thus, besides extending the earlier calculations already mentioned,
the present work makes contact and complements this recent work,
which is based on a treatment using the
kinetic equations of the neutrino-plasma system.
The results and formulas we present are applicable in a variety of
situations where the neutrino interactions with the other background
particles are important, and the method we employ could also be useful 
in the study of similar problems that may arise in other contexts.

A word about the strategy of our calculation is in order. In principle,
the diagrams involved are numerous, and many of them are shown
in \Ref{mohantynp}. However our calculation is simplified for
the following reason. Since we are interested in the terms that
contain both the neutrino and the electron distribution functions,
the momentum integrations are effectively cutoff. To order $1/M^2_W$, 
we can then replace the $W,Z$ boson propagators by their local limit,
so that the neutrino-electron interactions can be approximated by the
local Fermi interactions. In that limit, the set of diagrams that
are relevant to extract the contribution that we are seeking,
to order $1/M^2_W$, collapses to the class of diagram represented
in Fig. \ref{fig:pipenu}, which
we are considering in the manner we have indicated.

We begin, in Section\ \ref{sec:eprop} by writing down the expression for the
electron propagator that will be used in the calculation of the
photon self-energy, which includes the thermal effects as well
as the effects of the neutrino-electron interactions. In 
Section\ \ref{sec:selfenergy} the self-energy
tensor is calculated and in terms of the usual transverse and longitudinal
components and the two additional components that we have mentioned,
and the integral formulas for the latter are obtained and evaluated
for some cases in Section\ \ref{sec:formulas}. The photon dispersion
relations for this system are considered in Section\ \ref{sec:disprel},
the application to the study of the macroscopic
electromagnetic properties of this system is in Section\ \ref{sec:em},
and Section\ \ref{sec:conclusions} contains our conclusions.

%
%
\section{Electron propagator}
\label{sec:eprop}

The electron self-energy function $\Sigma_e(p)$ is determined
by computing the diagrams depicted in Fig.\ \ref{fig:sigmae}.
A straightforward calculation,
using the thermal propagator for the internal neutrino lines, 
yields to order $1/M^2_W$
\beq
\label{sigmae}
\Sigma_e = \lslash{v}(\lambda_V + \lambda_A\gamma_5) \,,
\eeq
where
\beqa
\label{lambdaVA}
\lambda_V & = & \sqrt{2}G_F\left[a^{(Z)}_e \sum_{\alpha = e,\mu,\tau} 
(n_{\nu_\alpha} - n_{\bar\nu_\alpha}) + n_{\nu_e} - n_{\bar\nu_e}\right]\,,
\nonumber\\*
\lambda_A & = & \sqrt{2}G_F\left[b^{(Z)}_e \sum_{\alpha = e,\mu,\tau} 
(n_{\nu_\alpha} - n_{\bar\nu_\alpha}) - 
(n_{\nu_e} - n_{\bar\nu_e})\right]\,.
\eeqa
Here the coefficients $a^{(Z)}_e, b^{(Z)}_e$ are the vector and axial
vector neutral-current couplings of the electron,
\beqa
a^{(Z)}_e & = & -\frac{1}{2} + 2\sin^2\theta_W \nonumber\\
b^{(Z)}_e & = & \frac{1}{2} \,,
\eeqa
and 
the $n_{\nu_\alpha},n_{\bar \nu_\alpha}$
denote respectively the total number density of each neutrino or antineutrino
specie in the medium. 

The electron propagator to be used in our calculation of the
photon self-energy is given by
\beq
\label{Se}
S_e(p) = S_F(p) - \left[S_F(p) - \bar S_F(p)\right]\eta_e(p\cdot u) \,,
\eeq
where $\bar S_F = \gamma^0 S^\dagger_F\gamma^0$, and
\beq 
\label{etae} 
\eta_e(p) = \theta(p\cdot u)f_e(p\cdot u) +
\theta(-p\cdot u)f_{\bar e}(-p\cdot u)\,, 
\eeq
with
\beqa 
f_e(x) & = & \frac{1}{e^{\beta(x - \mu_e)} + 1} \nonumber\\
f_{\bar e}(x) & = & \frac{1}{e^{\beta_e(x + \mu_e)} + 1} \,. 
\eeqa
Here $\beta_e$ is the inverse temperature and $\mu_e$ the
chemical potential of the electron gas.
In addition, $S_F$ is the electron propagator in the presence of
the neutrino gas, which is given by
\beq
\label{SF}
S^{-1}_F(p) = S^{-1}_0 - \Sigma_e(p)\,,
\eeq
where $S_0$ is the free propagator in the vacuum
\beq
\label{S0}
S_0 = \frac{\lslash{p} + m_e}{p^2 - m^2_e + i\epsilon} \,,
\eeq
and $\Sigma_e$ has been given in \Eq{sigmae}.
Therefore, to leading oder in $1/M^2_W$,
\beq
S_F(p) = S_0(p) + S_0(p)\Sigma_e(p)S_0(p)\,.
\eeq
Substituting this in \Eq{Se}, we obtain the electron thermal 
propagator to be used in our computation
\beq
\label{Secalc}
S_e = S_0 + S_T + S^\prime + S^\prime_T \,,
\eeq
where $S_0$ is given in \Eq{S0}, $S_T$ is the usual thermal
part of the electron propagator
\beq
S_T(p) = 2\pi i\delta(p^2 - m^2_e)\eta_e(p\cdot u)(\lslash{p} + m_e)\,,
\eeq
while
\beq
S^\prime(p) = \frac{(\lslash{p} + m_e)\Sigma_e(p)(\lslash{p} + m_e)}
{(p^2 - m^2_e + i\epsilon)^2} \,,
\eeq
and
\beq
\label{STprime}
S^\prime_T(p) = -2\pi i\delta^\prime(p^2 - m^2_e)\eta_e(p\cdot u)
(\lslash{p} + m_e)\Sigma_e(p)(\lslash{p} + m_e) \,.
\eeq
In \Eq{STprime}, $\delta^\prime$ denotes the derivative
of the delta function with respect to its argument.

In some respects, the present calculation resembles the calculation of the
photon self-energy in an electron background in the presence
of a magnetic field\cite{piBpal,piBdns}. 
In that case, the calculation involves
the use of the thermal generalization of the Schwinger
propagator, which takes into account the $B$ field and the thermal effects
of the electron background. In the present case, 
the neutrino current acts as the external field, playing the role
of the $B$ field in the former case.
%
%
\section{Photon self-energy}
\label{sec:selfenergy}

We decompose the photon self-energy into various parts according to whether
or not they depend on the neutrino densities. Therefore,
discarding the term that is independent of the particle densities,
we write
\beq
\label{pimunudef}
\pi_{\mu\nu} = \pi^{(e)}_{\mu\nu} + \pi^{(\nu)}_{\mu\nu} + 
\pi^{(e\nu)}_{\mu\nu}\,,
\eeq
where $\pi^{(e)}_{\mu\nu}$ is the purely electronic contribution, while
$\pi^{(\nu)}_{\mu\nu}$, which depends on the neutrino distributions
but not on the electron distribution, corresponds to the
contribution to the photon self-energy that was computed in \Ref{mohantynp}.
Although the results for $\pi^{(\nu)}_{\mu\nu}$, and of course
$\pi^{(e)}_{\mu\nu}$ are known, we state them below in the form 
that will be useful for later reference.
On the other hand, $\pi^{(e\nu)}_{\mu\nu}$
contains the terms that depend on both the neutrino and electron
distributions, and is the focus of the present paper.
To determine it, we start from the expression corresponding to the
diagram of Fig.\ \ref{fig:pipenu},
\beq 
\label{photonselfenergy} 
i\pi^\prime_{\mu\nu} \equiv
-(-ie)^2\int \frac{d^4p}{(2\pi)^4} \mbox{Tr}\,\left[\gamma_\mu
iS_e(p+q)\gamma_\nu iS_e(p)\right] \,,
\eeq
where $S_e(p)$ is given by \Eq{Se}, with $S_F$ determined from \Eq{SF}.
Substituting \Eq{Secalc} into \Eq{photonselfenergy} and singling out
the terms that contain the electron distributions as well as the 
neutrino densities we obtain,
\beqa
\label{pienudef}
i\pi^{(e\nu)}_{\mu\nu} & = & e^2\int\frac{d^4p}{(2\pi)^4}\left\{
\Tr\gamma_\mu iS_0(p^\prime)\gamma_\nu iS^\prime_T(p) +
\Tr\gamma_\mu iS^\prime_T(p^\prime)\gamma_\nu iS_0(p)\right.\nonumber\\
&&\mbox +
\left.\Tr\gamma_\mu iS^\prime(p^\prime)\gamma_\nu iS_T(p) +
\Tr\gamma_\mu iS_T(p^\prime)\gamma_\nu iS^\prime(p)\right\}\,,
\eeqa
where
\beq
p^\prime = p + q \,.
\eeq

\subsection{$\pi^{(e)}_{\mu\nu}$}
\label{sec:pieeval}

The pure electron term is given by
\beq
\label{piedef}
i\pi^{(e)}_{\mu\nu} = e^2\int\frac{d^4p}{(2\pi)^4}\left\{
\Tr\gamma_\mu iS_0(p^\prime)\gamma_\nu iS_T(p) +
\Tr\gamma_\mu iS_T(p^\prime)\gamma_\nu iS_0(p)\right\}\,.
\eeq
Computing the traces and carrying out the $p^0$ integration,
$\pi^{(e)}_{\mu\nu}$  can be written in the form
\beq
\label{pieexpr}
\pi^{(e)}_{\mu\nu} = -4e^2
\int\frac{d^3p}{(2\pi)^3 2E}(f_e + f_{\bar e})\left\{
\frac{L_{\mu\nu}}{q^2 + 2p\cdot q} + (q\rightarrow -q)\right\} \,,
\eeq
where
\beq
\label{Lmunu}
L_{\mu\nu} = 2p_\mu p_\nu + p_\mu q_\nu + q_\mu p_\nu - p\cdot q g_{\mu\nu} \,.
\eeq

The fact that $\pi^{(e)}_{\mu\nu}$ is symmetric and satisfies
$q^\mu \pi^{(e)}_{\mu\nu} = q^\nu \pi^{(e)}_{\mu\nu} = 0$ implies
that it is of the form
\beq
\label{piegen}
\pi^{(e)}_{\mu\nu} = \pi^{(e)}_T R_{\mu\nu} + \pi^{(e)}_L Q_{\mu\nu}\,,
\eeq
where
\beqa
\label{RQtensors}
R_{\mu\nu} & = & \tilde g_{\mu\nu} - Q_{\mu\nu}\,, \nonumber\\
Q_{\mu\nu} & = & \frac{\tilde u_\mu\tilde u_\nu}{\tilde u^2}\,,
\eeqa
with
\beq
\label{utilde}
\tilde u_\mu \equiv \tilde g_{\mu\nu}u^\nu \,,
\eeq
and
\beq
\tilde g_{\mu\nu} = g_{\mu\nu} - \frac{q_\mu q_\nu}{q^2} \,.
\eeq
In general $\pi^{(e)}_{T,L}$ are functions of the scalar variables
\beqa
\label{Qomega}
\omega & = & q\cdot u \nonumber\\
{Q} & = & \sqrt{\omega^2 - q^2} \,,
\eeqa
which have the interpretation of being the photon energy
and momentum, in the rest frame of the electron gas.  
The functions $\pi^{(e)}_{T,L}$ are determined by projecting 
\Eq{pieexpr} with the tensors $R_{\mu\nu}$ and $Q_{\mu\nu}$.  
This procedure then leads to
\beqa
\label{piTLe}
\pi^{(e)}_T & = & -2e^2\left(A_e
+ \frac{q^2}{{Q}^2}B_e\right) \,,\nonumber\\
\pi^{(e)}_L & = & 4e^2\frac{q^2}{{Q}^2}B_e
\eeqa
where
\beqa
\label{AeBe}
A_e & = & \int\frac{d^3{p}}{(2\pi)^3 2{E}}
(f_{e} + f_{\bar e})\left[
\frac{2m_e^2 - 2p\cdot q}{q^2 + 2p\cdot q} + 
(q\rightarrow -q)\right]\,,\nonumber\\
B_e & = & \int\frac{d^3{p}}{(2\pi)^3 2{E}}(f_{e} + f_{\bar e})
\left[
\frac{2(p\cdot u)^2 + 2(p\cdot u)(q\cdot u) - p\cdot q}
{q^2 + 2p\cdot q} + (q\rightarrow -q)\right]\,.
\eeqa
While these formulas can be used to evaluate the electronic contribution
to the photon self-energy in various situations, we need not proceed
any further in that direction since, 
as we have already mentioned, the results are well known and we can
simply quote the relevant ones when we need them. 

\subsection{$\pi^{(\nu)}_{\mu\nu}$}

In addition to the tensors $R_{\mu\nu}$ and $Q_{\mu\nu}$ that are
defined in \Eq{RQtensors}, we introduce 
\beqa
\label{Ptensors}
P_{\mu\nu} & = & \frac{i}{Q}\epsilon_{\mu\nu\alpha\beta} q^\alpha u^\beta\,,
\nonumber\\
P^\prime_{\mu\nu} & = & 
\frac{i}{Q}\epsilon_{\mu\nu\alpha\beta} q^\alpha v^{\prime\,\beta}\,,
\eeqa
with
\beq
v^{\prime\,\mu} = v^\mu - u^\mu (u\cdot v)\,.
\eeq
Then the results of \Ref{mohantynp} can be expressed in the form
\beq
\pi^{(\nu)}_{\mu\nu} = \pi^{(\nu)}_P (P^\prime_{\mu\nu} + 
u\cdot v P_{\mu\nu})\,,
\eeq
where 
\beq
\label{pipnu}
\pi^{(\nu)}_P = \frac{e^2 G_F}{\sqrt{2}\pi^2}\left(\frac{q^2}{m^2_e}\right)
Q(n_{\nu_e} - n_{\bar\nu_e})J(q^2) \,.
\eeq
$J(q^2)$ is given explicitly in that reference, but its precise value
will not be relevant here. We only wish to note the presence of
the kinematic factor of $q^2$ in \Eq{pipnu}. For the purpose of
determining the dispersion relations, this factor can be set
equal to the plasma frequency squared $\omega^2_p$, 
since in the lowest order $q^2 \sim \omega^2_p$.
However, in other applications, such the one as we will consider
in Section\ \ref{subsec:evolution}, the appropriate kinematic regime
corresponds to taking $q \rightarrow 0$, and in those cases
$\pi^{(\nu)}_P$ is not relevant.

\subsection{Evaluation of $\pi^{(e\nu)}_{\mu\nu}$}
Computing the traces and carrying out the $p^0$ integration in \Eq{pienudef},
$\pi^{(e\nu)}_{\mu\nu}$ can be expressed in the form
\beq
\label{pienumunu}
\pi^{(e\nu)}_{\mu\nu} = (-4e^2)(\lambda_VT^{(V)}_{\mu\nu} + 
\lambda_A T^{(A)}_{\mu\nu}) \,,
\eeq
where
\beqa
\label{TV}
T^{(V)}_{\mu\nu} & = & \int\frac{d^4p}{(2\pi)^3}\eta_e(p\cdot u)\left\{
\frac{-L^{(1)}_{\mu\nu}\delta^\prime(p^2 - m^2_e)}{d} +
\frac{L^{(2)}_{\mu\nu}\delta(p^2 - m^2_e)}{d^2} 
+ (q \rightarrow -q)\right\}
\nonumber\\
\label{TA}
T^{(A)}_{\mu\nu} & = & i\epsilon_{\mu\nu\alpha\beta}
\int\frac{d^4p}{(2\pi)^3}\eta_e(p\cdot u)\left\{
\frac{-K^{(1)\alpha\beta}\delta^\prime(p^2 - m^2_e)}{d} +
\frac{K^{(2)\alpha\beta}\delta(p^2 - m^2_e)}{d^2} 
- (q \rightarrow -q)\right\} \,,
\nonumber\\
\eeqa
with
\beqa
\label{LKmunu}
L^{(1)}_{\mu\nu} & = & (m^2_e - p^2)\left[p^\prime_\mu v_\nu - 
g_{\mu\nu}p^\prime\cdot v + v_\mu p^\prime_\nu\right] + 
(2p\cdot v)m^2_e g_{\mu\nu} +
2p\cdot v\left[p^\prime_\mu p_\nu + p_\mu p^\prime_\nu - 
p\cdot p^\prime g_{\mu\nu}\right] \nonumber\\
L^{(2)}_{\mu\nu} & = & (m^2_e - p^{\prime\,2})\left[p_\mu v_\nu - 
g_{\mu\nu}p\cdot v + v_\mu p_\nu\right] + 
(2p^\prime\cdot v)m^2_e g_{\mu\nu} +
2p^\prime\cdot v\left[p^\prime_\mu p_\nu + p_\mu p^\prime_\nu - 
p\cdot p^\prime g_{\mu\nu}\right] \nonumber\\
K^{(1)}_{\alpha\beta} & = & 
(p^2\, - m^2_e)p_{\alpha}v_\beta + (p^2 + m^2_e) q_{\alpha}v_\beta -
2p\cdot v\,q_{\alpha} p_\beta\,,
\nonumber\\
K^{(2)}_{\alpha\beta} & = & 
(m^2_e - p^{\prime\,2})p_{\alpha}v_\beta + 2m^2_e\, q_{\alpha}v_\beta
- 2p^\prime\cdot v\, q_{\alpha} p_\beta\,,
\eeqa
and
\beq
\label{d}
d = (p + q)^2 - m^2_e \,.
\eeq

\subsubsection{Evaluation of $T^{(V)}_{\mu\nu}$}
The integral expression for $T^{(V)}_{\mu\nu}$ can be simplified
as follows. Defining
\beqa
C_{\mu\nu} & = & p_\mu v_\nu + v_\mu p_\nu - p\cdot v g_{\mu\nu}\,,
\nonumber\\
D_{\mu\nu} & = & (m^2_e - p\cdot p^\prime)g_{\mu\nu} + 
p_\mu p^\prime_\nu + p^\prime_\mu p_\nu \,,
\eeqa
$L^{(1,2)}_{\mu\nu}$ can be expressed in the form
\beqa
L^{(1)}_{\mu\nu} & = & (m^2_e - p^2)
\left[v^\lambda\partial_\lambda D_{\mu\nu} -
C_{\mu\nu}\right] + 2p\cdot v D_{\mu\nu} \,,\nonumber\\
L^{(2)}_{\mu\nu} & = & -dC_{\mu\nu} + 2p^\prime\cdot v D_{\mu\nu} \,,
\eeqa
where $d$ is given in \Eq{d}, $\partial_\mu \equiv \partial/\partial p^\mu$,
and we have used the following relation,
\beq
p^\prime_\mu v_\nu + v_\mu p^\prime_\nu - p^\prime\cdot v g_{\mu\nu} =
v^\lambda\partial_\lambda D_{\mu\nu} - C_{\mu\nu} \,,
\eeq
which can be verified by explicit computation. Using the relations
\beqa
\label{derrelations}
(p^2 - m^2_e)\delta^\prime(p^2 - m^2_e) & = & -\delta(p^2 - m^2_e)\,,
\nonumber\\
v^\mu\partial_\mu\delta(p^2 - m^2_e) & = & 
2p\cdot v\delta^\prime(p^2 - m^2_e) \,,
\eeqa
it then follows that
\beq
\label{L1munu}
L^{(1)}_{\mu\nu}\delta^\prime(p^2 - m^2_e) =
v^\lambda\partial_\lambda\left[D_{\mu\nu}\delta(p^2 - m^2_e)\right] - 
C_{\mu\nu}\delta(p^2 - m^2_e)\,,
\eeq
and similarly
\beq
\label{L2munu}
\frac{L^{(2)}_{\mu\nu}}{d^2} = -D_{\mu\nu}v^\lambda\partial_\lambda
\left(\frac{1}{d}\right) - \frac{C_{\mu\nu}}{d} \,,
\eeq
where we have used
\beq
v^\lambda\partial_\lambda\left(\frac{1}{d}\right) = 
-2p^\prime\cdot v\frac{1}{d^2} \,.
\eeq
Therefore, from \Eq{TV}, with the help of \Eqs{L1munu}{L2munu},
\beq
T^{(V)}_{\mu\nu} =  \int\frac{d^4p}{(2\pi)^3}\eta_e(p\cdot u)\left\{
-v^\lambda\partial_\lambda\left[\frac{D_{\mu\nu}\delta(p^2 - m^2_e)}{d}\right] 
+ (q \rightarrow -q)\right\}\,,
\eeq
which, after a partial integration and then carrying out the integration
over $p^0$ using the delta function, yields
\beq
T^{(V)}_{\mu\nu} = v\cdot u \int\frac{d^3p}{(2\pi)^3 2 E}
\frac{\partial(f_e - f_{\bar e})}{\partial E}
\left\{\frac{L_{\mu\nu}}{q^2 + 2p\cdot q} + 
(q \rightarrow -q)\right\} \,,
\eeq
with $L_{\mu\nu}$ given in \Eq{Lmunu}. This has the same structure as
the normal electron background contribution, given in
\Eq{pieexpr}. Therefore, the same arguments used in Section\ \ref{sec:pieeval}
can be applied here to conclude that $T^{(V)}_{\mu\nu}$ can be expressed
in the form
\beq
\label{TVfinalexpr}
T^{(V)}_{\mu\nu} = \frac{1}{2}
\left(A^\prime_e + \frac{q^2}{Q^2}B^\prime_e\right) 
R_{\mu\nu} + \left(\frac{-q^2}{Q^2}\right)B^\prime_e Q_{\mu\nu} \,, 
\eeq
where
\beqa
\label{AeBeprime}
A^\prime_e & = & u\cdot v\int\frac{d^3{p}}{(2\pi)^3 2{E}}
\frac{\partial(f_{e} + f_{\bar e})}{\partial E}\left[
\frac{2m_e^2 - 2p\cdot q}{q^2 + 2p\cdot q} + 
(q\rightarrow -q)\right]\,,\nonumber\\
B^\prime_e & = & u\cdot v\int\frac{d^3{p}}{(2\pi)^3 2{E}}
\frac{\partial(f_{e} + f_{\bar e})}{\partial E}
\left[
\frac{2(p\cdot u)^2 + 2(p\cdot u)(q\cdot u) - p\cdot q}
{q^2 + 2p\cdot q} + (q\rightarrow -q)\right]\,.
\eeqa

\subsubsection{Evaluation of $T^{(A)}_{\mu\nu}$}
From \Eq{LKmunu}, using \Eq{derrelations} we can write
\beq
\frac{-K^{(1)}_{\alpha\beta}\delta^\prime(p^2 - m^2_e)}{d} +
\frac{K^{(2)}_{\alpha\beta}\delta(p^2 - m^2_e)}{d^2} = 
2m^2_e q_\alpha v_\beta \left[\frac{\delta(p^2 - m^2_e)}{d^2} -
\frac{\delta^\prime(p^2 - m^2_e)}{d}\right] + v^\lambda\partial_\lambda\left(
\frac{q_\alpha p_\beta\delta(p^2 - m^2_e)}{d}\right) \,.
\eeq
Therefore,
\beq
\label{TAaux}
T^{(A)}_{\mu\nu} = i\epsilon_{\mu\nu\alpha\beta}q^\alpha\left[I_1^\beta  +
I_2 v^\beta\right] \,,
\eeq
where
\beqa
\label{KI1}
I^\beta_1 & = & \int\frac{d^4p}{(2\pi)^3)}\eta_e(p\cdot u)\left\{
v^\lambda\partial_\lambda\left[\frac{p^\beta\delta(p^2 - m^2_e)}{d}\right]
\right\} + (q\rightarrow -q)\,,\nonumber\\
\label{KI2}
I_2 & = & 2m^2_e\int\frac{d^4p}{(2\pi)^3)}\eta_e(p\cdot u)\left\{
\frac{\delta(p^2 - m^2_e)}{d^2} -
\frac{\delta^\prime(p^2 - m^2_e)}{d}\right\} + (q\rightarrow -q) \,.
\eeqa
For $I^\beta_1$, by partial integration we obtain
\beq
\label{I1beta}
I^\beta_1 = -v\cdot u\int\frac{d^3p}{(2\pi)^3 2E}
\frac{\partial(f_e + f_{\bar e})}{\partial E} p^\beta
\left[\frac{1}{q^2 + 2p\cdot q} + (q \rightarrow -q)\right] \,.
\eeq
Since the integral is a function only of the vectors $q^\mu$ and $u^\mu$,
it must be of the form
\beq
I^\beta_1 = I_1 u^\beta + I^\prime_1 q^\beta \,.
\eeq
Substituting it in \Eq{TAaux},
\beq
\label{TAfinalexpr}
T^{(A)}_{\mu\nu} = i\epsilon_{\mu\nu\alpha\beta}q^\alpha\left[I_1 u^\beta +
I_2 v^\beta\right] \,,
\eeq
and therefore $I^\prime_1$ is not relevant. On the other hand,
$I_1$ can be determined by using the projection relation
\beq
\tilde u_\beta I^\beta_1 = \tilde u^2 I_1 \,,
\eeq
which, from \Eq{I1beta}, yields the formula
\beq
\label{I1intformula}
I_1 = \frac{v\cdot u}{Q^2}\int\frac{d^3p}{(2\pi)^3 2E}
\frac{\partial(f_e + f_{\bar e})}{\partial E}
\left[\frac{q^2(p\cdot u) - (q\cdot u)(p\cdot q)}{q^2 + 2p\cdot q}\right] + 
(q \rightarrow -q)\,.
\eeq

To carry out the integration over $p^0$ in \Eq{KI2} we write
\beq
\delta^\prime(p^2 - m^2_e) = \frac{1}{2p^0}\frac{\partial}{\partial p^0}
\delta(p^2 - m^2_e) \,,
\eeq
and using the usual rule for the integration over the derivative of
the delta function after some algebra we obtain
\beq
\label{I2intformula}
I_2 = 2m^2_e\int\frac{d^3p}{(2\pi)^3 2E}\frac{\partial}{\partial E}\left[
\frac{1}{2E}\left(\frac{f_e + f_{\bar e}}{q^2 + 2p\cdot q}\right)\right] 
+ (q\rightarrow -q)\,.
\eeq

\subsection{Summary}

From the results we have obtained, it follows that
the photon self-energy can be expressed in the form
\beq
\label{pitotalexpr}
\pi_{\mu\nu} = \pi_T R_{\mu\nu} + \pi_L Q_{\mu\nu} + 
\pi_{P} P_{\mu\nu} + \pi^\prime_{P}P^\prime_{\mu\nu}\,,
\eeq
where $P_{\mu\nu}$ and $P^\prime_{\mu\nu}$ have been defined in \Eq{Ptensors}.
Using \Eqs{piegen}{pienumunu}, together with the results 
for $T^{(V,A)}_{\mu\nu}$ given in \Eqs{TVfinalexpr}{TAfinalexpr}, 
the coefficients in \Eq{pitotalexpr} are then given by
\beqa
\label{picoefftotal}
\pi_T & = & \pi^{(e)}_T -
2e^2\lambda_V\left(A^\prime_e + \frac{q^2}{Q^2}B^\prime_e\right)\,,\nonumber\\
\pi_L & = & \pi^{(e)}_L + 
4e^2\lambda_V\left(\frac{q^2}{Q^2}\right)B^\prime_e\,,\nonumber\\
\pi_P & = & (u\cdot v)\pi^{(\nu)}_P + \pi^{(e\nu)}_P \nonumber\\
\pi^\prime_P & = & \pi^{(\nu)}_P + \pi^{\prime(e\nu)}_P \nonumber\\
\eeqa
where
\beqa
\label{piPformulas}
\pi^{(e\nu)}_{P} & = & -4e^2 \lambda_A Q[I_1 + (u\cdot v) I_2]\,,\nonumber\\
\pi^{\prime(e\nu)}_{P} & = & -4e^2 \lambda_A Q I_2 \,,
\eeqa
with $I_{1,2}$ given by \Eqs{I1intformula}{I2intformula}, respectively. 

The terms proportional to $\lambda_V$ give only small correction to
the pure electronic terms. For example, consider the case of a classical
electron distribution, for which we can use
\beq
\label{dfeclass}
\frac{\partial f_{e,\bar e}}{\partial E} \approx -\beta f_{e,\bar e} \,.
\eeq
Then from \Eq{AeBeprime},
\beqa
A^\prime_e & = & -\beta A_e\,,\nonumber\\
B^\prime_e & = & -\beta B_e\,,
\eeqa
and remembering \Eq{piTLe}, it follows that the neutrino-dependent contribution
to $\pi_{T,L}$ is smaller than the electron term by a factor of 
$\beta\lambda_V$. Assuming, for illustrative purposes, that the neutrino
gas can also be treated classically, so that $n_\nu \sim T^3$, then
$\beta\lambda_V \sim G_F T^2$, which is negligible in most situations
of interest. Therefore, for all practical purposes of interest to us,
we will neglect that contribution in \Eq{picoefftotal} and
use
\beqa
\pi_T & = & \pi^{(e)}_T\,,\nonumber\\
\pi_L & = & \pi^{(e)}_L
\eeqa
in what follows.

%
%
\section{Evaluation of $\pi^{(e\nu)}_P$ and $\pi^{\prime(e\nu)}_P$}
\label{sec:formulas}

Useful formulas for $\pi^{(e\nu)}_{P}$ and $\pi^{\prime(e\nu)}_P$
can be obtained by evaluating the
integrals $I_{1,2}$ defined in \Eqs{I1intformula}{I2intformula} in the
long wavelength limit. The expressions for $I_{1,2}$ in that
limit can be obtained by applying the auxiliary 
formula\cite{nucharge1,nucharge2}
\beq
\label{I12lwaux}
\int\frac{d^3p}{(2\pi)^3}\frac{{\cal F}(p,q)}{[q^2 + 2p\cdot q]^n} = 
\int\frac{d^3p}{(2\pi)^3}\frac{\left({\cal F} - 
\frac{\vec Q}{2}\cdot\frac{d{\cal F}}{d\vec p} - 
\frac{n\omega}{2E}{\cal F}\right)}
{[2E\omega - 2\vec p\cdot\vec Q]^n} \,,
\eeq
and they are valid for
\beq
\label{lwlimit}
\omega, Q \ll\langle{\cal E}\rangle\,,
\eeq
where $\langle{\cal E}\rangle$ denotes a 
typical average energy of the electrons in the background. 
In \Eq{I12lwaux} the symbol $\frac{d}{d\vec p}$ stands for
the total momentum derivative,
\beq
\label{pder}
\frac{d}{d\vec p} = \frac{\partial}{\partial\vec p} + \frac{\vec p}{E}
\frac{\partial}{\partial E} \,.
\eeq

Let us consider $I_2$ first. We rewrite \Eq{I2intformula} in the form
\beq
I_2 = \frac{m^2_e}{2}\int\frac{d^3p}{(2\pi)^3}\left\{
\frac{f^\prime}{q^2 + 2q\cdot p} - \frac{2\omega f}{[q^2 + 2q\cdot p]^2}
\right\} + (q \rightarrow -q)\,,
\eeq
where, to simplify the notation, we have defined
\beqa
f & = & \frac{f_e + f_{\bar e}}{E^2} \,,\nonumber\\
f^\prime & = & \frac{1}{E}\frac{\partial}{\partial E}\left(
\frac{f_e + f_{\bar e}}{E}\right) \,.
\eeqa
Then by direct application of \Eq{I12lwaux} we obtain
\beqa
\label{I2lw}
I_2 & = & \frac{m^2_e}{4}\int\frac{d^3p}{(2\pi)^3}
\frac{\omega}{E[\omega E - \vec p\cdot\vec Q]^2}
\left[\vec p\cdot\vec Q\frac{\partial f}{\partial E} + 2\omega f \right]
\nonumber\\
&&\mbox{} - \frac{m^2_e}{4}\int\frac{d^3p}{(2\pi)^3}
\frac{1}{E[\omega E - \vec p\cdot\vec Q]}
\left[\vec p\cdot\vec Q\frac{\partial f^\prime}{\partial E} + 
\omega f^\prime \right] \,.
\eeqa
Furthermore, denoting by $\bar v$ the average velocity of the particles
in the background, the following approximate formulas
are useful for practical applications,
\beq
\label{I2lwapproximate}
I_2 = \left\{
\begin{array}{ll}
\frac{m^2_e}{4}\int\frac{d^3p}{(2\pi)^3}
\frac{1}{E}\frac{\partial}{\partial E}\left[
\frac{1}{E}\frac{\partial}{\partial E}\left(\frac{f_e + f_{\bar e}}{E}\right)
\right] & (\omega \ll \bar v Q)\\[12pt]
-\frac{m^2_e}{4}\int\frac{d^3p}{(2\pi)^3}
\frac{1}{E}\frac{\partial}{\partial E}\left(\frac{f_e + f_{\bar e}}{E^3}\right)
& (\omega \gg \bar v Q) \,,
\end{array}\right. 
\eeq
which are obtained from \Eq{I2lw} by taking the static limit $\omega = 0$, 
or the $\vec Q = 0$ limit, respectively.
The remaining integration can be carried out simply for particular cases of
the distribution functions, and we consider some of them below.

Regarding $I_1$, it can be written in the form
\beq
I_1 = \frac{u\cdot v}{2Q^2}\int\frac{d^3p}{(2\pi)^3}
\frac{F_1}{q^2 + 2p\cdot q}
+ (q\rightarrow -q) \,,
\eeq
where
\beq
F_1 \equiv \frac{\partial(f_e + f_{\bar e})}{\partial E}\left[
-Q^2 + \frac{\omega\vec p\cdot\vec Q}{E}\right] \,,
\eeq
and applying \Eq{I12lwaux},
\beq
\label{I1aux}
I_1 = -\frac{u\cdot v}{4Q^2}\int\frac{d^3p}{(2\pi)^3}\left(
\vec Q\cdot \frac{dF_1}{d\vec p} + \frac{\omega}{E}F_1\right)
\frac{1}{E\omega - \vec p\cdot\vec Q}\,.
\eeq
To compute the momentum derivative we use
\beq
\frac{\partial F_1}{\partial E} = 
-Q^2\frac{\partial^2(f_e + f_{\bar e})}{\partial E^2} + 
\omega\vec p\cdot\vec Q\frac{\partial}{\partial E}\left[
\frac{1}{E}\frac{\partial(f_e + f_{\bar e})}{\partial E}\right]
\eeq
which, using $\vec p\cdot\vec Q = E\omega - (E\omega - \vec p\cdot\vec Q)$,
we write as
\beq
\frac{\partial F_1}{\partial E} = 
-Q^2\frac{\partial^2(f_e + f_{\bar e})}{\partial E^2} + 
\omega^2 E\frac{\partial}{\partial E}\left[
\frac{1}{E}\frac{\partial(f_e + f_{\bar e})}{\partial E}\right] -
\omega(E\omega - \vec p\cdot\vec Q)\frac{\partial}{\partial E}\left[
\frac{1}{E}\frac{\partial(f_e + f_{\bar e})}{\partial E}\right]\,.
\eeq
In addition
\beq
\frac{\partial F_1}{\partial\vec p} = \frac{\omega}{E}
\frac{\partial(f_e + f_{\bar e})}{\partial E}\vec Q \,,
\eeq
and therefore, remembering \Eq{pder},
\beq
\frac{\omega}{E}F_1 + \vec Q\cdot\frac{dF_1}{d\vec p} = 
q^2\frac{\vec p\cdot\vec Q}{E}\frac{\partial^2(f_e + f_{\bar e})}{\partial E^2}
- \omega\frac{\vec p\cdot\vec Q}{E}
(E\omega - \vec p\cdot\vec Q)\frac{\partial}{\partial E}\left[
\frac{1}{E}\frac{\partial(f_e + f_{\bar e})}{\partial E}\right]\,.
\eeq
When this is substituted in \Eq{I1aux}, the second term gives zero
by symmetric integration, and we obtain
\beq
\label{I1lw}
I_1(\omega,Q) = -\frac{u\cdot v}{4}\frac{q^2}{Q^2}
\int\frac{d^3p}{(2\pi)^3}\frac{1}{E}
\frac{\partial^2(f_e + f_{\bar e})}{\partial E^2}\frac{\vec p\cdot\vec Q}
{E\omega - \vec p\cdot Q} \,.
\eeq
To evaluate $I_1$ in the $\vec Q \rightarrow 0$ limit, we first
expand the denominator in the integrand,
\beq
\frac{1}{E\omega - \vec p\cdot\vec Q} = \frac{1}{E\omega}\left[
1 + \frac{\vec p\cdot\vec Q}{E\omega}\right]\,.
\eeq
The term proportional to a single power of $\vec p\cdot\vec Q$ integrates
to zero, while in the quadratic term we can put 
$(\vec p\cdot\vec Q)^2\rightarrow \frac{1}{3}p^2 Q^2$. In this way,
in analogy with \Eq{I2lwapproximate}, we obtain
the approximate formulas
\beq
\label{I1lwapproximate}
I_1 = \left\{
\begin{array}{ll}
-\frac{u\cdot v}{4}\int\frac{d^3p}{(2\pi)^3}\frac{1}{E}
\frac{\partial^2(f_e + f_{\bar e})}{\partial E^2} & 
(\omega \ll \bar v Q)\\[12pt]
-\frac{u\cdot v}{4}\int\frac{d^3p}{(2\pi)^3}
\frac{p^2}{3E^3}\frac{\partial^2(f_e + f_{\bar e})}{\partial E^2} &
(\omega \gg \bar v Q)\,.
\end{array}\right.
\eeq
In what follows we give explicit results for several useful cases.

\subsection{Non-relativistic, non-degenerate gas}

For a classical and non-relativistic gas, using \Eq{dfeclass}
and $f_{\bar e} \simeq 0$, \Eqs{I2lwapproximate}{I1lwapproximate} yield 
\beq
I_1 = -u\cdot v\,I_2\,,
\eeq
with
\beq
I_2 = 
n_e\times\left\{
\begin{array}{ll}
\frac{\beta^2}{8m_e} & (\omega \ll \bar v Q)\\[12pt]
\frac{\beta}{8m^2_e} & (\omega \gg \bar v Q)\,,
\end{array}\right.
\eeq
where $n_e$ is the total number density of electrons.

\subsection{Relativistic, non-degenerate gas}

In this case we also assume for simplicity that $f_{\bar e} \simeq f_e$. Then,
for $I_1$, we use \Eq{dfeclass} twice, and in the limit $m_e \rightarrow 0$
\beq
I_1 = -\frac{u\cdot v}{4\pi^2}\times
\left\{
\begin{array}{ll}
1 & (\omega \ll \bar v Q)\\[12pt]
\frac{1}{3} & (\omega \gg \bar v Q)\,.
\end{array}\right.
\eeq
For $I_2$ we have to be careful because the integral
is not defined for $m_e = 0$. As we show below, the result is
\beq
\label{I2erclass}
I_2 = \frac{1}{4\pi^2}\,,
\eeq
independently of whether $\omega \ll \bar v Q$ or $\omega \gg \bar v Q$.
To illustrate how we have proceeded 
let us consider the $\omega \gg \bar v Q$ case in some detail.
By carrying out the angular integration, followed by an integration
by parts, from \Eq{I2lwapproximate} we obtain the formula
\beq
\label{I2erclassaux}
I_2 = \frac{m^2_e}{8\pi^2}\int^\infty_0 dp\,\frac{(f_e + f_{\bar e})}{E^3} \,.
\eeq
By making the change of variable $p = m_e\tan\theta$, 
\beqa
\int dp\,\frac{f_e + f_{\bar e}}{E^3} & = & \frac{1}{m^2_e}\int d\theta\,
(f_e + f_{\bar e})\cos\theta\nonumber\\
& = & \frac{2}{m^2_e}\int d\theta\, e^{-\beta m\sec\theta}\cos\theta \,,
\eeqa
which we evaluate by using the Taylor expansion of the
exponential. In this way we obtain
\beq
\int^\infty_0 dp\frac{f_e + f_{\bar e}}{E^3} = \frac{2}{m^2_e}[1 + O(m_e)] \,,
\eeq
and substituting it in \Eq{I2erclassaux} we arrive at \Eq{I2erclass}.
In similar fashion, for $\omega \ll \bar v Q$,
\beq
I_2 = -\frac{m^2_e}{8\pi^2}\int dp\,\frac{1}{E}\frac{\partial}{\partial E}
\left(\frac{f_e + f_{\bar e}}{E}\right)\,.
\eeq
Using \Eq{dfeclass} and then making the change of variables as above,
we obtain
\beq
\int dp\,\frac{1}{E}\frac{\partial}{\partial E}
\left(\frac{f_e + f_{\bar e}}{E}\right) = 
-\frac{2}{m^2_e}\left[1 + O(m_e)\right]\,,
\eeq
which leads to \Eq{I2erclass}.

\subsection{Degenerate gas}

For a degenerate gas, whether it is relativistic or not,
\beqa
f_e & = & \theta(E_F - E) \,,\nonumber\\
f_{\bar e} & \simeq & 0\,,
\eeqa
where $E_F = \sqrt{p^2_F + m^2_e}$ is the Fermi energy, with
\beq
p_F = (3\pi^2 n_e)^{1/3}
\eeq
being the Fermi momentum. Then we obtain in this case
\beqa
I_1 & = & -\left(\frac{u\cdot v}{8\pi^2}\right)\times
\left\{
\begin{array}{ll}
\frac{E_F}{p_F} & (\omega \ll \bar v Q)\\[12pt]
\frac{p_F}{E_F}\left[1 - \frac{2p^2_F}{3E^2_F}\right] 
& (\omega \gg \bar v Q)\,,
\end{array}
\right.
\nonumber\\[12pt]
I_2 & = & \left(\frac{1}{8\pi^2}\right)\times
\left\{
\begin{array}{ll}
\frac{E_F}{p_F} & (\omega \ll \bar v Q)\\[12pt]
\frac{p_F}{E_F} & (\omega \gg \bar v Q)\,.
\end{array}
\right.
\eeqa

Furthermore, in the non-relativistic (NR) or
the extremely relativistic (ER) case, this yields
\beq
I^{(NR)}_1 = -u\cdot v I^{(NR)}_2 = 
-\left(\frac{u\cdot v}{8\pi^2}\right)\times
\left\{
\begin{array}{ll}
\frac{m_e}{p_F} & (\omega \ll \bar v Q)\\[12pt]
\frac{p_F}{m_e} & (\omega \gg \bar v Q)\,,
\end{array}
\right.
\eeq
\beq
I^{(ER)}_1 = -\left(\frac{u\cdot v}{8\pi^2}\right)
\left\{
\begin{array}{ll}
1 & (\omega \ll \bar v Q)\\[12pt]
\frac{1}{3} & (\omega \gg \bar v Q)\,,
\end{array}
\right.
\eeq
and
\beq
I_2^{(ER)} = \frac{1}{8\pi^2}
\eeq
independently of whether $\omega \ll \bar v Q$ or $\omega \gg \bar v Q$.

\subsection{Explicit formulas for $\pi^{(e\nu)}_P$ and $\pi^{\prime(e\nu)}_P$}
\label{subsec:pipformulas}

The formulas that we have obtained for $I_{1,2}$ allow us to
evaluate $\pi^{(e\nu)}_P$ and $\pi^{\prime(e\nu)}_P$ for typical situations
of interest in physical applications.
We specifically consider the regime 
\beq
\omega \gg \bar v Q\,,
\eeq
which will be of special interest to us in Section\ \ref{subsec:evolution}.
Thus, for example, if the electron gas is degenerate,
\beqa
\pi^{(e\nu)}_P & = & -\left(\frac{e^2}{3\pi^2}\right)
\frac{\lambda_A p^3_F}{E^3_F}Q\,u\cdot v\,,
\nonumber\\
\pi^{\prime(e\nu)}_P & = & -\left(\frac{e^2}{2\pi^2}\right)
\frac{\lambda_A p_F}{E_F}Q \,.
\eeqa
These expressions hold whether the electrons are relativistic or not.
On the other hand, for a classical and relativistic electron gas,
\beqa
\pi^{(e\nu)}_P & = & -\left(\frac{2e^2}{3\pi^2}\right)\lambda_A
Q\,u\cdot v\,,\nonumber\\
\pi^{\prime(e\nu)}_P & = & -\left(\frac{e^2}{\pi^2}\right)\lambda_A Q \,.
\eeqa

Regarding the value of $\lambda_A$, from \Eq{lambdaVA},
\beq
\label{lambdaAstd}
\lambda_A=\frac{G_F}{\sqrt{2}} 
\left [(n_{\nu_{\tau}}-n_{\bar\nu_{\tau}} ) + 
(n_{\nu_{\mu}}-n_{\bar\nu_{\mu}} ) - 
(n_{\nu_{e}}-n_{\bar\nu_{e}})\right]\,.
\eeq
Denoting by $\mu_{\nu_x}$ the chemical potential of the neutrino of 
flavor $x = e,\mu,\tau$, the approximate formulas
\beq
\label{lambdaAhighT}
\lambda_A = \frac{G_F}{6\sqrt{2}} T^3 
\left ( \xi_{\nu_{\tau}}+\xi_{\nu_{\mu}}-
\xi_{\nu_{e}}\right) \qquad (\mu_{\nu_x}\ll T)
\eeq
or
\beq
\label{lambdaAhighmu}
\lambda_A = \frac{G_F}{6\sqrt{2}\pi^2} 
\left ( \mu^3_{\nu_{\tau}}+\mu^3_{\nu_{\mu}}-\mu^3_{\nu_{e}}\right )
\qquad (\mu_{\nu_x}\gg T) \,,
\eeq
can be useful.

\subsection{Discussion}

It should be noted that the contributions
$\pi^{(e\nu)}_P$ and $\pi^{\prime(e\nu)}_P$ that we have calculated
in this work, have a very different kinematic dependence on the photon
momentum $q^\mu$ if we compare them with the term $\pi^{(\nu)}_P$
that was calculated in \Ref{mohantynp} and the analogous
quantity calculated in \Ref{repko}. In particular, 
$\pi^{(\nu)}_P$, which does not depend explicitly on the electron distribution,
is proportional to $q^2$, as we have indicated in \Eq{pipnu}. 
For the purpose of determining the
photon dispersion relations, since $q^2$ is of the order of
the plasma frequency squared, the value of $\pi^{(\nu)}_P$ turns out to
be proportional to the electron density and in fact comparable to
the values of $\pi^{(e\nu)}_P$ and $\pi^{\prime(e\nu)}_P$.
In the absence of the electrons, $\pi^{(e\nu)}_P$ and $\pi^{\prime(e\nu)}_P$
are of course zero and, since $q^2 \simeq 0$, $\pi^{(\nu)}_P$ is negligible.
Thus, in a pure neutrino gas the term calculated in \Ref{repko},
although it is of higher order in $1/M^2_W$, is the dominant one.
On the other hand, in applications such as the one 
that we consider in Section\ \ref{subsec:evolution}, in which
the relevant kinematic limit is $q\rightarrow 0$, corresponding
to the long wavelength and static regime,
$\pi^{(\nu)}_P$ does not contribute and 
$\pi^{(e\nu)}_P$ and $\pi^{\prime(e\nu)}_P$ are the only relevant ones.

%
%
\section{Dispersion relations}
\label{sec:disprel}

In the presence of an external current $j^{(ext)}_\mu$, 
the electromagnetic potential
in the medium is determined from the field equation, which
in momentum space is
\beq
\label{claseqmotion}
\left[-q^2\tilde g_{\mu\nu} + \pi_{\mu\nu}\right] A^\nu = j^{(ext)}_\mu \,.
\eeq
The photon dispersion relations are determined by finding the solutions
of the homogeneous equation, i.e., with $j^{(ext)}_\mu = 0$.
To specify the various modes it is convenient to introduce the basis
vectors $\epsilon^\mu_{1,2,3}$, which we defined as follows. 

For a given photon momentum $\vec Q$,
we define the unit vectors $\hat e_i$ ($i = 1,2,3$) by writing
\beq
\vec Q \equiv Q \hat e_3 \,,
\eeq
with $\hat e_{1,2}$ chosen such that
\beqa
\hat e_1\cdot \hat e_3 & = & e_1\cdot \hat e_3 = 0 \,,\nonumber\\
\hat e_2 & = & \hat e_3 \times \hat e_1 \,.
\eeqa
For the problem that we are considering,
without loss of generality, we can choose the vectors
$\hat e_{1,2}$ such that $\vec V$ lies in
the $1,3$ plane. Thus, the unit vector
\beq
\hat V = \frac{\vec V}{V} \,,
\eeq
has the decomposition
\beq
\hat V = \cos\theta\,\hat e_3 + \sin\theta\,\hat e_1 \,.
\eeq
where
\beq
\cos\theta = \hat Q\cdot\hat V \,.
\eeq
The vectors $\epsilon^\mu_{1,2,3}$ are then defined by
\beqa
\epsilon^\mu_1 & = & (0, \hat e_1) \,,\nonumber\\
\epsilon^\mu_2 & = & (0, \hat e_2) \,,\nonumber\\
\epsilon^\mu_3 & = & \frac{1}{\sqrt{q^2}}(Q, \omega\hat e_3) \,,
\eeqa
which form a basis of vectors orthogonal to $q^\mu$. They 
satisfy the relations
\beqa
\label{RQPmatrixelements}
R_{\mu\nu}\epsilon^\nu_3 = P_{\mu\nu}\epsilon^\nu_3 = 
Q_{\mu\nu}\epsilon^\nu_{1,2} & = & 0\,,\nonumber\\
R_{\mu\nu}\epsilon^\nu_i & = & \epsilon_{i\mu}\,, \qquad \mbox{($i = 1,2$)}
\,,\nonumber\\
Q_{\mu\nu}\epsilon^\nu_3 & = &  \epsilon_{3\nu}\,,\nonumber\\
P_{\mu\nu}\epsilon^\nu_1 & = & i\epsilon_{2 \mu} \,,\nonumber\\
P_{\mu\nu}\epsilon^\nu_2 & = & -i\epsilon_{1 \mu} \,,
\eeqa
from which we can immediately read off the matrix elements
of the tensors $R,Q,P$ between any pair of the basis 
vectors $\epsilon^\mu_{1,2,3}$. In addition, the only non-zero
matrix elements of the tensor $P^\prime$ are given by
\beqa
\label{Pprimematrixelements}
\epsilon^\mu_2 \epsilon^\nu_3 P^\prime_{\mu\nu} =
-\epsilon^\mu_3 \epsilon^\nu_2 P^\prime_{\mu\nu} & = & 
-\frac{i\sqrt{q^2}}{Q}V\sin\theta \,,\nonumber\\
\epsilon^\mu_1 \epsilon^\nu_2 P^\prime_{\mu\nu} =
-\epsilon^\mu_2 \epsilon^\nu_1 P^\prime_{\mu\nu} & = &
-\frac{i\omega}{Q}V\cos\theta \,.
\eeqa

To find the propagating modes, we express the polarization
vectors in the form
\beq
\xi^\mu = \sum^3_{i = 1} \alpha_i \epsilon^\mu_i \,,
\eeq
where the coefficients $\alpha_i$ and the corresponding 
dispersion relations are to be found by solving the equation
\beq
\label{classeqhom}
\left[-q^2\tilde g_{\mu\nu} + \pi_{\mu\nu}\right] \xi^\nu = 0 \,.
\eeq
With the help of the relations given 
in \Eqs{RQPmatrixelements}{Pprimematrixelements},
this equation can be written in matrix notation
\beq
(q^2 - \Pi)\mathbf{\alpha} = 0 \,,
\eeq
where
\beq
\mathbf{\alpha} = \left(\begin{array}{c}\alpha_1\\ \alpha_2\\
\alpha_3\end{array}\right) \,, 
\eeq
and
\beq 
\label{pimatrix} 
\Pi = \left(\begin{array}{ccc}
\pi_T & -i\pi_P + i\frac{\omega}{Q}V\pi^\prime_P\cos\theta & 0 \\
i\pi_P - i\frac{\omega}{Q}V\pi^\prime_P\cos\theta & \pi_T &
i\frac{\sqrt{q^2}}{Q}V\pi^\prime_P\sin\theta \\
0 & -i\frac{\sqrt{q^2}}{Q}V\pi^\prime_P\sin\theta & \pi_L
\end{array}\right) \,.
\eeq
In what follows we consider some particular cases 
of this equation, whose solutions reveal the structure and the main features
of more general ones.

\subsection{$\vec V = 0$ case}

In this case the $\pi^\prime_P$ term in \Eq{pitotalexpr} is absent, 
and consequently the dispersion
relations have the same form as those studied earlier in Ref.\ \cite{nppip}. 
The longitudinal mode has the dispersion relation
\beq
\omega^2 - Q^2 - \pi_L = 0\,,
\eeq
and corresponding polarization vector
\beq
\label{xiL}
\xi^\mu_L = \epsilon^\mu_3 \,.
\eeq
The polarization vectors of the transverse modes are then given by
\beq
\label{xipm}
\xi^\mu_{\pm} = \frac{1}{\sqrt{2}}(\epsilon^\mu_1 \pm i\epsilon^\mu_2) \,,
\eeq
with the corresponding dispersion relations given by
\beq
\omega^2 - Q^2 - (\pi_T \pm \pi_P) = 0 \,,
\eeq
respectively.

\subsection{$\vec V \not= 0$ case}

The presence of the $\pi^\prime_P$ term in \Eq{pitotalexpr} 
in general modifies the picture described above, and can give rise
to new effects. This is due to the fact that the 
existence of the velocity vector $\vec V$ of the neutrino gas
breaks the three-dimensional isotropy of the medium, for example
not too differently from the way in which it would
be broken by the presence of an external magnetic field.
We consider two special situations.

\subsubsection{Propagation parallel to $\vec V$}

In this case $\sin\theta = 0$, and
\beq 
\label{pimatrixparallel} 
\Pi = \left(\begin{array}{ccc}
\pi_T & -i\pi_P + i\frac{\omega}{Q}V\pi^\prime_P & 0 \\
i\pi_P - i\frac{\omega}{Q}V\pi^\prime_P & \pi_T & 0 \\
0 & 0 & \pi_L
\end{array}\right) \,.
\eeq
Therefore, the longitudinal mode is unaffected, while for the 
transverse modes the polarization vectors are the same as in \Eq{xipm}
but the dispersion relations are now given by
\beqa
\omega^2 - Q^2 -
\left(\pi_T + \pi_P - \frac{\omega}{Q}V\pi^\prime_P\right) & = & 0\,,
\nonumber\\
\omega^2 - Q^2  - 
\left(\pi_T - \pi_P + \frac{\omega}{Q}V\pi^\prime_P\right) & = & 0\,,
\eeqa
for $\xi_\pm$, respectively.

\subsubsection{Propagation perpendicular to $\vec V$}

In this case, setting $\theta = \pi/2$, 
\beq 
\label{pimatrixperp} 
\Pi = \left(\begin{array}{ccc}
\pi_T & -i\pi_P  & 0 \\
i\pi_P & \pi_T & i\frac{\sqrt{q^2}}{Q}V\pi^\prime_P \\
0 & -i\frac{\sqrt{q^2}}{Q}V\pi^\prime_P & \pi_L
\end{array}\right) \,.
\eeq
Therefore the modes are neither purely longitudinal nor
transverse to $\vec Q$. Finding the general solution in this case is a 
cumbersome process. In some circumstances it may be appropriate to 
seek approximate formulas for the dispersion relations and polarization vectors
as a perturbative expansion in $V\pi^\prime_P$, and that 
can be extended to other values of $\theta$ as well.

Some applications of the optical activity effects
induced by neutrinos were considered in \Ref{mohantynp}. 
As a rule, the effects tend to be small.
Our main intention in this section was to indicate how the photon
dispersion relations are modified by the anisotropic effects produced
by a non-zero velocity of the neutrino gas relative to the electron
background. It can be of interest to carry this further to study the 
implications of these effects in the context of the specific applications 
considered in \Ref{mohantynp}, or similar ones. However,
that is outside the scope and focus of the applications that we
have considered, to which we now turn our attention.

%
%
\section{Macroscopic Electrodynamics}
\label{sec:em}

Besides modifying the dispersion relations of the propagating modes,
the presence of the neutrino gas influence the
electromagnetic properties of the system in the
static and long wavelength regime. To study them, it is
useful to formulate the results or our calculations 
using the language of macroscopic electrodynamics.  

In what follows we will 
assume that the neutrino gas is at rest with respect to the electron gas,
that is
\beq
v^\mu = u^\mu \,,
\eeq
since this case already brings out the essential consequences of the presence
of the neutrino gas. In this case, $P^\prime_{\mu\nu} = 0$,
and the photon self-energy takes the form
\beq
\pi_{\mu\nu} = \pi^{(e)}_T R_{\mu\nu} + \pi^{(e)}_L Q_{\mu\nu} + 
\pi_{P}P_{\mu\nu} \,.
\eeq
As has been discussed previously\cite{nppip,nppipE}, this is indeed the
most general of the photon self-energy in an isotropic medium,
which is the case if $\vec V = 0$.

\subsection{Dielectric function}
\label{sec:dielectric}

Introducing the electromagnetic field 
\beq
F_{\mu\nu} = -i(q_\mu A_\nu - A_\mu q_\nu)\,,
\eeq
the equation of motion, \Eq{claseqmotion}, can be written in the form
\beq
\label{maxwell}
-iq^\mu F_{\mu\nu} = j^{(ext)}_\nu + j^{(ind)}_\nu\,,
\eeq
where
\beq
\label{jinddef}
j^{(ind)}_\mu = -\pi_{\mu\nu}A^\nu \,.
\eeq
In fact, \Eq{maxwell} is equivalent to the Maxwell equations, with
$j^{(ind)}_\mu$ interpreted as the induced current. For example,
take the component of \Eq{maxwell} corresponding to the index $\nu$ being
a spatial index. With the usual definition of the fields,
\beqa
\label{EBdef}
\vec E & = & i\omega\vec A - i\vec Q A^0 \,,\nonumber\\
\vec B & = & i\vec Q\times\vec A \,,
\eeqa
which are related by
\beq
\label{EBrel}
\vec B = \frac{1}{\omega}\vec Q\times\vec E\,,
\eeq
the equation is just
\beq
\label{maxwellB}
i\vec Q\times\vec B + i\omega\vec E = \vec j^{(ext)} + \vec j^{(ind)} \,.
\eeq
Moreover, using \Eq{EBdef} in \Eq{jinddef}, the induced current
is given in terms of the fields by
\beq
\label{jindclass}
\vec \jmath^{\,(ind)} = i\omega\left[(1 - \epsilon_l)\vec E_l
+ (1 - \epsilon_t)\vec E_t + i\epsilon_p \hat{Q}\times\vec E\right] \,,
\eeq
where
\beqa
\label{epsilonpirel}
1 - \epsilon_t & = & \frac{\pi_T}{\omega^2}\,, \nonumber\\
1 - \epsilon_l & = & \frac{\pi_L}{q^2}\,,\nonumber\\
\epsilon_p & = & \frac{\pi_P}{\omega^2} \,,
\eeqa
and the longitudinal and transverse components of the electric field are
defined by  
\beqa
\label{Etl}
\vec E_l & = & \hat{Q}(\hat{Q}\cdot\vec E) \,,\nonumber\\
\vec E_t & = & \vec E - \vec E_l \,.
\eeqa
\Eq{jindclass} is the most general form of the induced current, 
involving terms that are linear in the field, 
and subject only to the assumption of isotropy.  
The quantities $\epsilon_{t,l}$ in \Eq{epsilonpirel} are transverse and 
longitudinal components of the dielectric constant of the
medium. Alternatively, instead of $\epsilon_{t,l}$, the dielectric and
magnetic permeability functions $\epsilon,\mu$ are
introduced by writing the induced current in the equivalent form
\beq
\label{jindclassequiv}
\vec \jmath^{(ind)} = i\left[\omega(1 - \epsilon)\vec E
+ \left(1 - \frac{1}{\mu}\right)\vec Q\times \vec B
+ i\frac{\omega^2}{Q} \epsilon_p\vec B\right] \,,
\eeq
where we have used \Eq{EBrel}.
Comparing \Eqs{jindclass}{jindclassequiv}, the relations
\beqa
\label{epsilonmu}
\epsilon & = & \epsilon_l\,,\nonumber\\
\frac{1}{\mu} & = & 1 + \frac{\omega^2}{{Q}^2}(\epsilon_l - \epsilon_t) \,,
\eeqa
then follow. 

These equations can of course be used to discuss the dispersion
relations of the propagating modes and related effects. However
we do not proceed any further in this direction since that
would essentially reproduce what we have already considered in 
Section\ \ref{sec:disprel}. Instead we turn our attention to
another kind of effect that can arise due to the presence
of the neutrino gas, which can be described on the basis
of these equations together with the results we have obtained.

\subsection{Evolution of magnetic fields}
\label{subsec:evolution}

Here we consider the evolution of an initial magnetic perturbation
in this medium. In the absence of any external sources, 
the equation for the magnetic field, \Eq{maxwellB}, is
\beq
i\vec Q\times\vec B + i\omega\vec E = \vec j^{(ind)} \,,
\eeq
which using \Eq{jindclassequiv} we can write in the form
\beq
\epsilon\omega\vec E + \frac{1}{\mu}\vec Q\times\vec B +
i\gamma\vec B = 0 \,,
\eeq
with
\beq
\label{gammadef}
\gamma = -\frac{\omega^2}{Q}\epsilon_p  = -\frac{\pi_P}{Q}\,.
\eeq
Since we are interested in following the evolution of $\vec B$,
we eliminate $\vec E$ from this equation by taking the cross product
with $\vec Q$ and then using \Eq{EBrel}, which yields
\beq
\label{evBmomentum}
\epsilon\omega^2\vec B - \frac{1}{\mu}Q^2\vec B + 
i\gamma\vec Q\times\vec B = 0\,.
\eeq
We obtain the corresponding equation in coordinate space by taking
the long wavelength limit, $\omega \gg \bar v Q$, and making the
quasi-static approximation, $\omega \rightarrow 0$. In this limit 
\beq
\epsilon \rightarrow  1 + \frac{i\sigma}{\omega} \,,
\eeq
where $\sigma$ is the conductivity of the medium. Therefore
the equation becomes
\beq
i\omega\sigma\vec B - \frac{1}{\mu}Q^2\vec B + 
i\gamma\vec Q\times\vec B = 0\,,
\eeq
or, in coordinate space,
\beq
\label{Beq}
\sigma\frac{\partial\vec B}{\partial t} = \frac{1}{\mu}\nabla^2\vec B +
\gamma\vec\nabla\times\vec B\,.
\eeq

Although we are using the same symbols $\mu$ and $\gamma$,
in \Eq{Beq} they stand for the corresponding quantities
evaluated in the long wavelength and static limit, as indicated above.
In this limit, $\pi^{(\nu)}_P$ does not contribute to $\gamma$
in \Eq{gammadef}, as can be seen from \Eq{pipnu}. Then 
using the results for $\pi^{(e\nu)}_P$ summarized in
Section\ref{subsec:pipformulas}, we can readily determine
$\gamma$ for some particular cases. For example,
for the relativistic and non degenerate electron gas,
\beq
\gamma = \frac{2e^2\lambda_A}{3\pi^2}\,,
\eeq
while for a degenerate electron gas,
\beq
\gamma = \frac{e^2\lambda_A}{3\pi^2}\left(\frac{p_F}{E_F}\right)^3 \,,
\eeq
with $\lambda_A$ given in \Eq{lambdaAstd}.
For more general cases, $\gamma$ can be computed
from the formula
\beq
\gamma = -e^2\lambda_A\left[\int\frac{d^3p}{(2\pi)^3}
\frac{p^2}{3E^3}\frac{\partial^2(f_e + f_{\bar e})}{\partial E^2} +
m^2_e \int\frac{d^3p}{(2\pi)^3}
\frac{1}{E}\frac{\partial}{\partial E}\left(\frac{f_e + f_{\bar e}}{E^3}\right)
\right]\,,
\eeq
which follows from the expression for $\pi^{(e\nu)}_P$ given in
\Eq{piPformulas}, together with the formulas for $I_{1,2}$ given
in \Eqs{I2lwapproximate}{I1lwapproximate} for $\omega \gg \bar v Q$.

An equation of the same form as \Eq{Beq} was obtained by
Semikoz and Sokoloff\cite{semikozalpha} by very different means,
in their work suggesting a new mechanism for the generation
of large-scale magnetic fields in the Early Universe
as a consequence of the neutrino-plasma interactions.
As emphasized in that reference, the equation describes the self-excitation
of an almost constant magnetic perturbation. 

Our approach is useful in two complementary ways. 
On one hand, it sheds light on the 
physical origin of this mechanism. Within our formulation, it is
a consequence of the optical activity induced by the interaction of
neutrinos with the other background particles. On the other hand, it
puts this mechanism on a firm footing from a computational point of view.
The optical activity induced by the neutrinos can be characterized
by the presence of the $\pi_P$ (and $\pi^\prime_P$) terms
in the photon self-energy, which also has a definite interpretation
in terms of the components of the dielectric function that enter
in the macroscopic (Maxwell) equations in the medium. Therefore,
by focusing on these quantities, we are able to give well defined
formulas for the parameters that are relevant to this effect,
in a way that are applicable to a variety 
of astrophysical and cosmological situations in which
the presence of neutrinos is influential. Our work also paves the way
to incorporate some corrections that can be important in specific
applications, such as the anisotropic effects that can arise
if the neutrino gas has a non-zero velocity relative to the
electron background. Further studies along these lines
are the subject of current work.

%
%
\section{Conclusions}
\label{sec:conclusions}

We have studied the electromagnetic properties of
an electron background, that contains a neutrino gas which 
is either at rest or moving as a whole relative to the background.
Apart from the well known longitudinal and transverse polarization functions
of the photon in a medium,
the presence of the neutrinos gives rise to two additional 
polarization functions, that we denote by $\pi_P$ and $\pi^{\prime}_P$.
We have computed that particular contribution to these two functions,
$\pi^{(e\nu)}_P$ and $\pi^{\prime(e\nu)}_P$,
that depends on the neutrino-antineutrino asymmetry in the medium 
as well as the momentum integral of the
electron (and positron) distribution function. The integrals
were evaluated for various specific conditions of the electron gas, 
and explicit formulas that are useful in many situations were given.
One of the consequences of
a non-zero value of $\pi_P$ and $\pi^{\prime}_P$ is to give rise to 
birefringence and anisotropic effects in the propagation of a photon through
that medium. We analyzed various particular situations to indicate
how the anisotropies due to the non-zero velocity of the neutrino
gas can affect the optical activity of the system.
A non-zero value of $\pi_P$ and $\pi^{\prime}_P$ also has
consequences related to the electromagnetic properties of the
system at a macroscopic level, and we considered specifically 
the evolution of a macroscopic magnetic field in this system. 
We arrived at an equation for the dynamics of the magnetic field,
that had been suggested in \Ref{semikozalpha} as a mechanism for
the generation of large-scale magnetic fields in the Early Universe
as a consequence of the neutrino-plasma interactions. In this
way we established contact between our work and this particular 
kind of application, which has been of recent interest. 
The approach we followed, which has been based on the application of
Finite Temperature Field Theory, as well as the calculations and results 
that are presented here, helps to put this subject on a firm footing
and to set a basis for carrying out further studies and applications along
these lines using powerful calculational techniques.

\begin{acknowledgments}

This material is based upon work supported by the US National
Science Foundation under Grant No. 0139538 (JFN); and by
DGAPA-UNAM under PAPIIT grant number IN119405 (SS).

\end{acknowledgments}

\end{document}